\documentclass{article}
\usepackage{sao2,psfig}
\issue{2000, 49, 45-52}
\begin{document}
\title{An algorithm for cleaning extended  interferences in radio astronomical records}
\author{O.V. Verkhodanov, D.A. Pavlov }
\institute{\saoname}
\date{February 17, 2000}{March 9, 2000}
\maketitle

\begin{abstract}
An algorithm is described for removing extended interferences, for instance
from a radar, which are shorter than the time of passage of a radio source
across the beam of the radio telescope. The algorithm is developed on the
basis of robust procedures with the application of extreme statistics.
A procedure of removing ``jumps'' in observational records is also described.
The results of performance of the algorithms are illustrated in the figures.
The cleaning procedures are included in the standard data reduction system
of RATAN-600.
\end{abstract}

\section{Introduction}
In the course of astrophysical observations one often encounters
interferences that corrupt the arriving useful signal or the statistical
noise estimates in observation records. This is especially pronounced
in records of radio astronomical observations. In the processing of data,
a number of noises are recorded that require special cleaning, which
is frequently done manually, for instance, in the standard reduction system
FADPS at RATAN-600 (Verkhodanov et al., 1993).

To perform the tasks of record cleaning from discontinuous jamming,
extensive use is made of technical means at the detector level (see
Berlin et al., 1995). However, when working with the archival data,
one has to come across different kinds of noises that disturb the initial
record. Such noises
may be ordinary discontinuous jamming, radar signals, and also various
jumps caused by the instability of ADC in data recording.

Recent work concerned with the development of the archive ODA (Kononov
et al., 1998; Kononov, Pavlov, 1999; Kononov et al., 1999) of RATAN-600
observational data coming from the broad-band radiometers made it possible
to solve the problems which demand the use of old data.
Among such problems one may isolate, for instance, investigation of the
background radiation (Parijskij, Korol'kov, 1986) and  search for
variability of radio astronomical objects. The processing of these
data has naturally called for automatic search and removal of
interferences that corrupt the signal. Some work over control of noises
by programme means based on nonparametric evaluation of the mean was
done at RATAN-600 several years ago (Erukhimov et al., 1990;
Chernenkov, 1996; Shergin et al., 1997). One should also  mention
 the programme development based on a gradient approach of removing
discontinuous jamming which was initiated by Bursov (1987). The work
over the creation of the cleaning procedure described below was
started by Pavlov in 1997 (Pavlov, 1998) under the supervision of the
former of the authors of this paper, when the problem of search and
removal of solitary interference from the radar was resolved.

It is proposed in this paper to develop an algorithm capable of
removing non-single interferences on a record. The algorithm employs both
the robust (immune to influence) estimates of the mean on the basis
of non-parametric rank algorithms (Huber, 1981) and the procedures based
on extreme statistics, namely, minimax  approach (Verkhodanov,
Gorokhov, 1995), which enable fast evaluation of statistics changes in
a given interval.

Below is presented a detailed description of removal of two classes of
interferences: 1) radar and discontinuous jamming; 2) jumps in records.

\begin{figure*}
\centerline{\psfig{figure=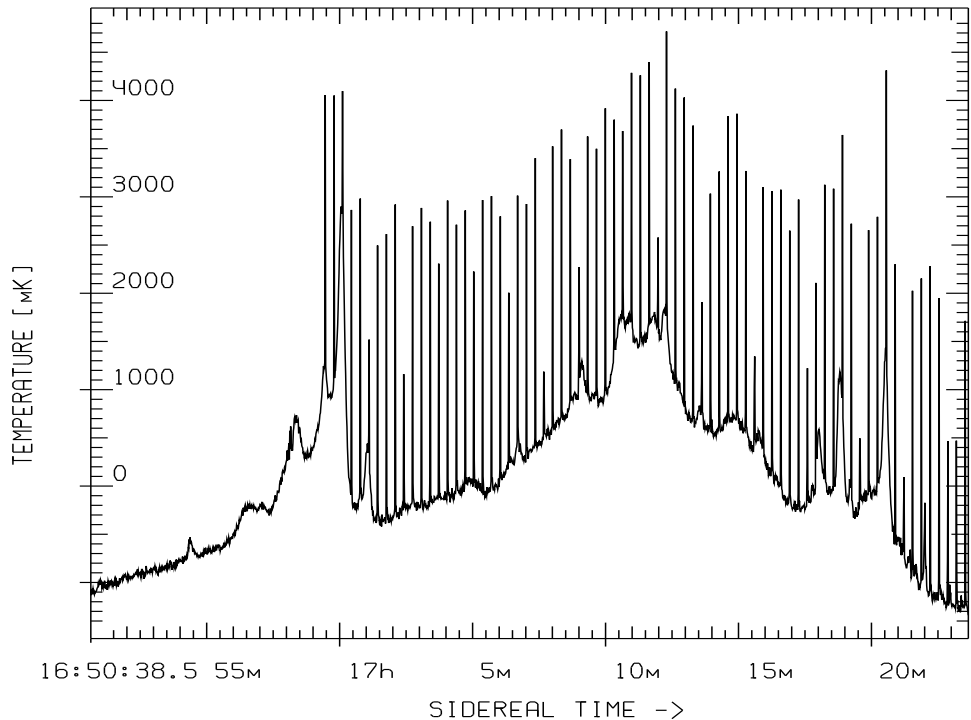,width=14cm,height=10cm,angle=-90}}
\caption{Example of RATAN-600 observations at 13~cm on 28/08/1994
with  the locator
operated. The record of the Galactic Plane is covered with  the ``forest''
of interferences (courtesy of S.A.Trushkin).}
\end{figure*}

\section{Algorithm of radar interference cleaning}
The problems presented by the signal from a radar (Fig. 1) are that it is not
a $\delta$-function, that is, it cannot be detected with the aid of
the ``gradient'' approach without confusion with a real source, and it
is quasiperiodic, i.e. it is ``fuzzed up'' in the region of spatial
frequencies, which hampers its removal on the Fourier plane. Besides,
because the signal is extended, the clipping (namely, removal of all points
exceeding a specified level with subsequent replacement by pixels
according to given rules) of its spatial spectrum will result in
corruption of the signal. For this reason the authors decided to use
statistical techniques to detect and remove interferences produced by
the radar. The developed procedure permitted getting rid of simple
discontinuous jamming. In principle, within the frame of FADPS one can
establish a procedure of removing interferences from radars on the
basis of the already available modules (Verkhodanov et al., 1992). For
instance, such an algorithm can be synthesized from the modules of low
noise subtraction, subtraction of the signal of Gaussian shape with a
size larger than the specified one, clipping the record at a given
level, and taking the sum of the record made with the subtracted
low-frequency noises and detected sources. However, because of the
multiplicity of Gaussians being computed and also the repetition of vector
operations of addition/subtraction increasing the calculation time of the script
operation, and due to the ambiguities arising in superposition of a
useful signal and a radar signal, the authors made a decision to abandon
this approach.

Prior to describing of the developed algorithm we will
explain some terms being employed in this paper. By the window is meant
the interval on the abscissa axis (for our data this is a time interval
measured in seconds) inside of which a signal search is made. By
sliding the window along the record we mean the sequential filling of
the array of pixels in the appropriate interval when the window is shifted by
1 pixel along the record. By the robust evaluation of the mean is implied
a simple median of distribution of pixel values in the given interval,
while the robust dispersion is the dispersion estimate made via the
median absolute deviation (MAD) under the assumption of Gaussian noises
(in this case if $\mu$ is MAD, then $\sigma$=$\mu$/0.674 in accordance
with the definition of second-order moments for the normal distribution).

When developing the algorithm, we used the archival data (Kononov et al.,
1999) written in the form of the FITS-like F format (Verkhodanov et al., 1993)
of the data processing system FADPS at RATAN-600.

The algorithm eliminating interferences from the radar comprises several
steps:
\begin{enumerate}
\item Copying of data from the F file to the working array and evaluation of the
robust dispersion $\sigma$ of the total record.\\
\item Search for minimum and maximum values in the window sliding along
the one-dimensional array. The window size is chosen equal to the beam
of the radio telescope. As the window is sliding, an array of maximum
peak-to-peak difference is filled in the current interval. The maximum
scatter is computed as the difference of the maximum and minimum values
of the pixels ({\it max--min}) in the current limits along the axis of shift.\\
\item When looking for  the maximum among the values of differences, which
exceeds the level $n \cdot \sigma$, where $n$ is the assigned level (by
default 3), analysis is made to search for the maximum locations by way
of simple comparing the array values.\\
\item In the region where the local maximum is found
the behaviour of the statistics of the surrounding pixels is
determined in case the signal does not fall fully within the window.
With this aim in view, the current position of the interval with respect
to the signal being detected is found: right and left of the maximum,
by means of robust averaging of neighbouring points. Depending on the
current vicinity (right or left) of the observed signal, the location
of the interval in which the parameters will be estimated is chosen.
The real maximum must always fall within the given interval.\\
\item Once the window position has been found, statistical estimate of
principal characteristics of the detected signal is made. For this
purpose the following formulae (Fomalont, 1989) are used:
   $$
       I  = \sum_{t} I_{t}
   $$
   $$
       X  = \frac{1}{I}\sum_{t} X_{t} I_{t}   \eqno (6)
   $$
   $$
       B  = \sqrt {\frac{1}{I} (X_{t} - X)^{2} I_{t} }, \qquad t=1,r;
   $$

where $I_t$ is the intensity value of the $t$-th point in the current
interval, $X_t$ is the location of this point in the record, $X$ is
the coordinate of the source centre of gravity, $I$ is the integral intensity
of a possible sought-for signal, $B$ is the object size, $r$ is the size
of the surveyed interval in pixels.\\
\item Comparison of the found parameters with the specified ones to
search for an interfering signal. If the determined amplitude is smaller than
the specified one or the size (duration) of the signal exceeds the given
one (i.e. a real source is detected), the window is then shifted by one
pixel and a return to the point of calculation of ({\it max--min}) statistics
is accomplished.\\
\item If the parameters correspond to the expected properties of an
interference, the boundaries of the signal to be removed are defined.\\
\item In the procedure of replacement of the points of the record within
the limits found, provision is made for variants of replacing the
generated noise with characteristic properties (dispersion and mean)
of the noise process of the present record or ordinary regression by two
estimated means at the edges of the record bordering on the current interval.
To generate a noise
Gaussian process, the standard library function of the language ``C'' {\it
rand()} is used, which outputs uniformly distributed values. For
conversion to the Gaussian noise, we applied a simple procedure of averaging
with a shifted mean and normalizing:
$$
N_i = \sigma\sqrt{12m}\frac{2\sum\limits^m_i R - m}{2m},
$$

where $R$ is the vector of random numbers of the uniform distribution
of the function {\it rand()} returned by the generator, $m$ is the number
of the points involved in averaging of uniformly distributed numbers. The
larger $m$, the better the distribution corresponds to normal. We
employed $m$=32 in our procedure. The value of the current system time is
used as the origin of the generator.\\
\item If there is one more maximum in a specified interval, then return
to the point of search for the location of the maximum in the current window
(see above) is performed.\\
\item If the end of the record is not reached, a shift is made by the
size of the interval of cleaning along the axis.\\
\item When the record is finished the data are output to the F file.
\end{enumerate}

\begin{figure*}
\centerline{\psfig{figure=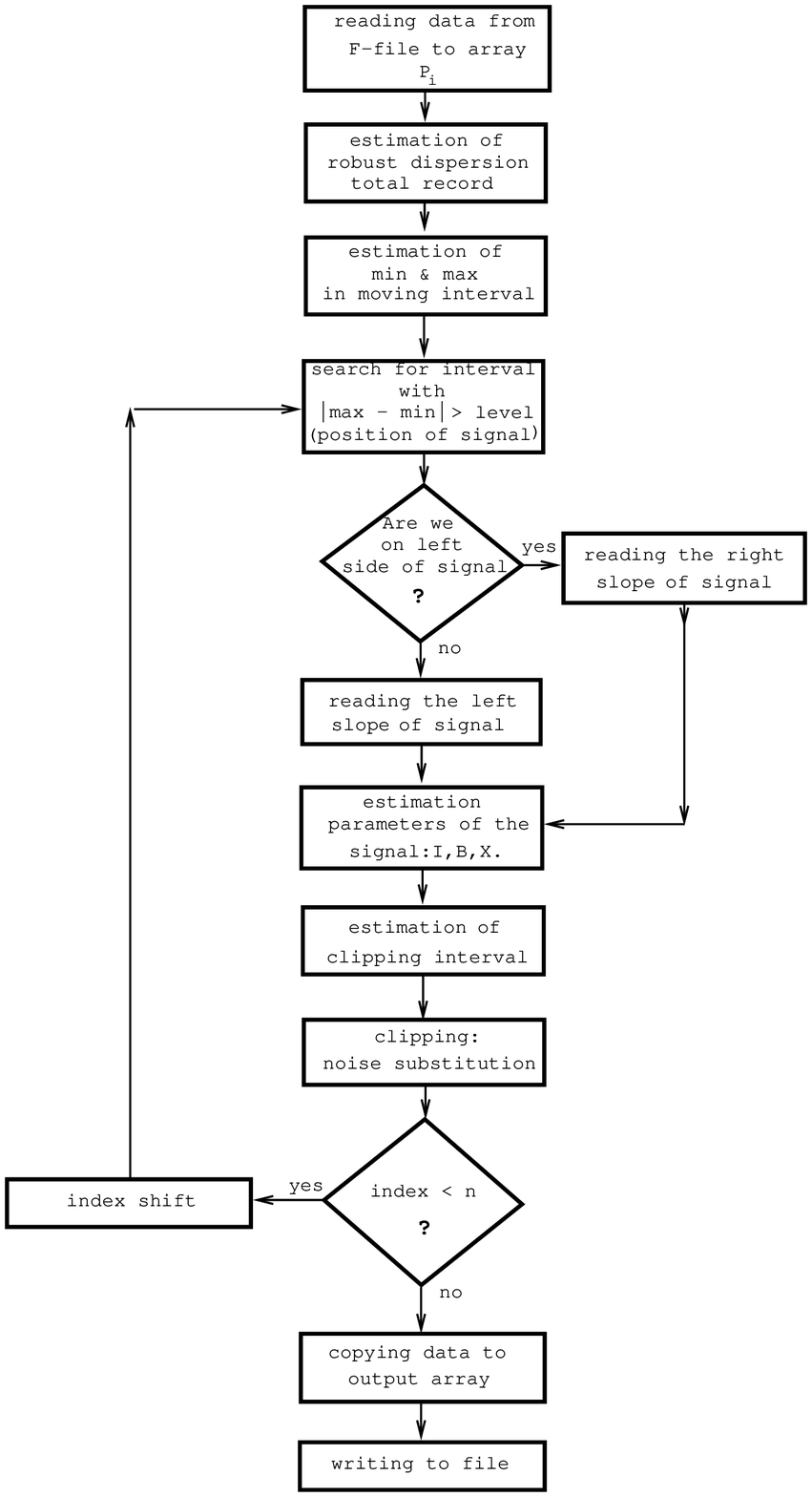,width=14cm,height=24cm}}
\caption{Block-diagram of search for and clipping  point-like and extended
interferences.}
\end{figure*}

The schematic of the radar interference removal algorithm described
above is displayed in Fig.2.

The results of operation of the algorithm presented in Fig.3.

Fig.4 shows the power Fourier spectra of the two records displayed in
Fig.3: prior to cleaning and after it. To highlight the difference, one
and the same curve of the low-frequency background is subtracted from
both records. The difference between the spectra is caused by elimination
of the discontinuous jamming and quasiperiodic radar signal.

\begin{figure*}
\centerline{\psfig{figure=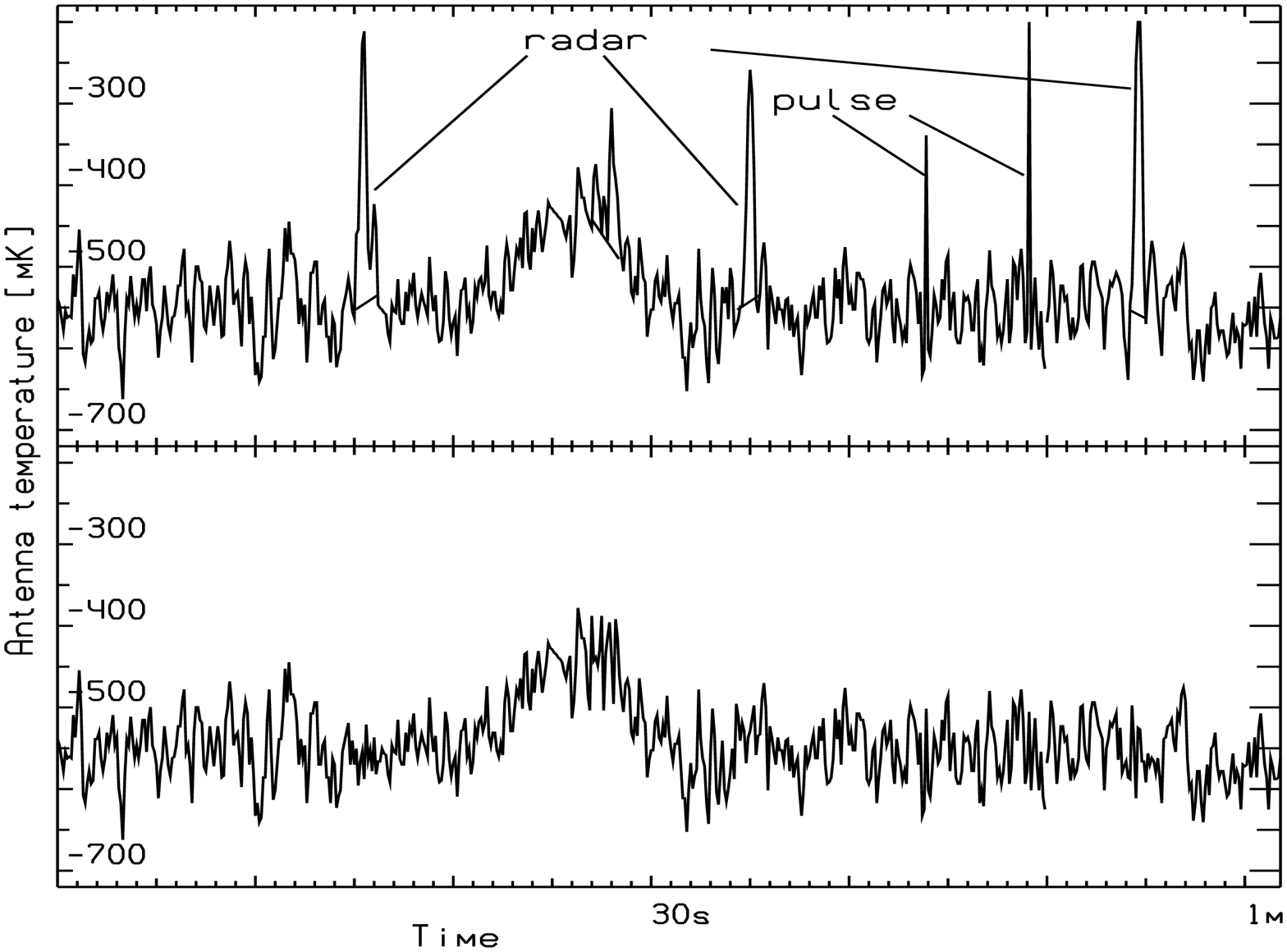,width=14cm,height=10cm,angle=-90}}
\caption{Results of clipping interferences in a record of wide-band
radiometers. The upper panel shows a record with a source and impulse, and
locator interferences. The lower panel shows a  record cleaned with the
algorithm of extended interference clipping with noise addition.}
\end{figure*}
\begin{figure*}
\centerline{\psfig{figure=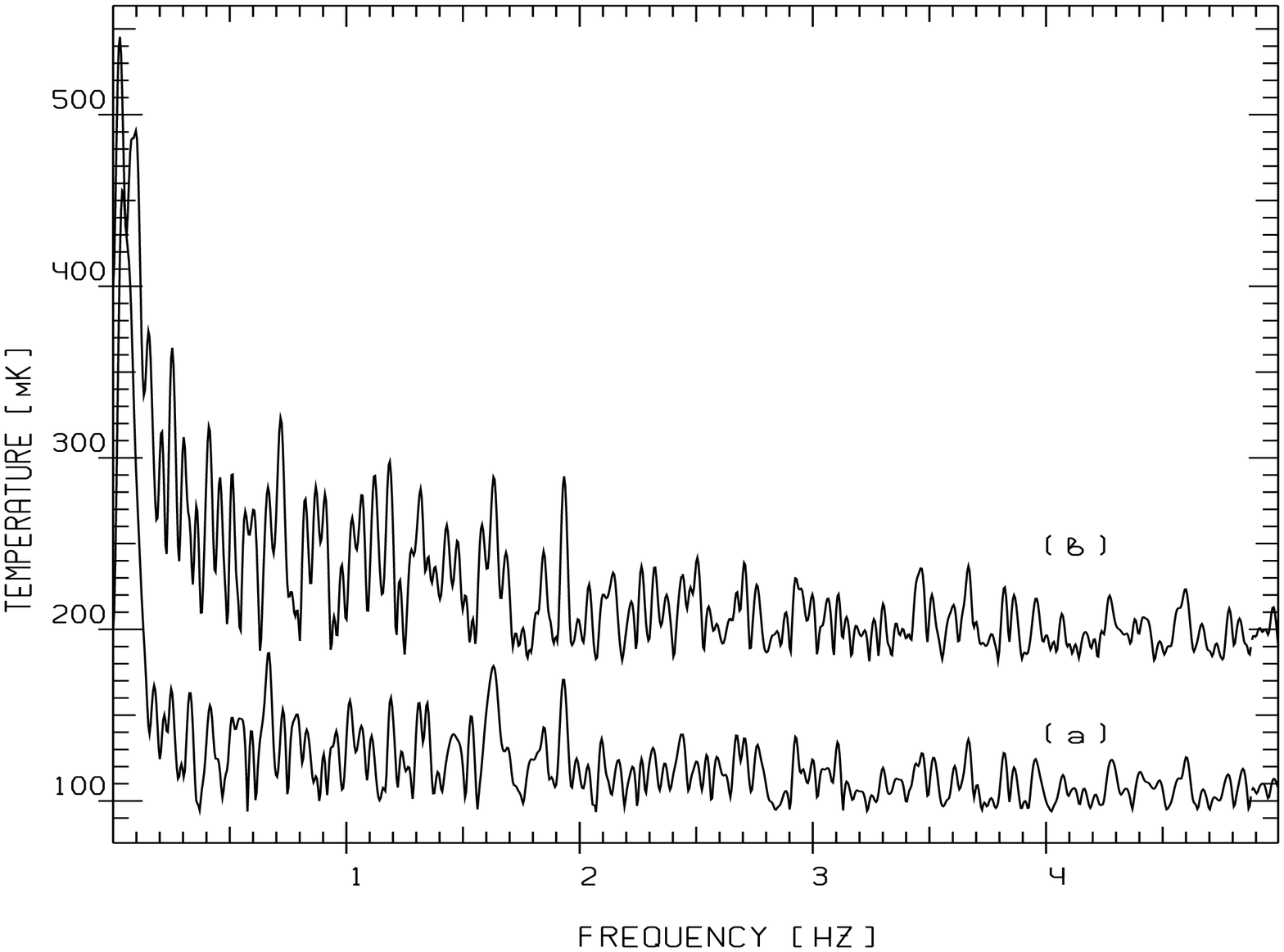,width=14cm,height=10cm,angle=-90}}
\caption{Power Fourier spectrum of records before (a) and after (b) cleaning
(see Fig. 3). The difference is due to cleaning of two impulse
interferences and a quasiperiodic signal of the locator.}
\end{figure*}

\section{Algorithms of removing ``jumps''}
Apart from interferences treated by the described algorithm, there is
a class of interferences not infrequently found in archival records:
jumps of the mean or ``steps'' (Fig.6). We have developed a procedure
of search for the jump on the basis of analysis of the mean over four
neighbouring sliding intervals.

\begin{figure*}
\centerline{\psfig{figure=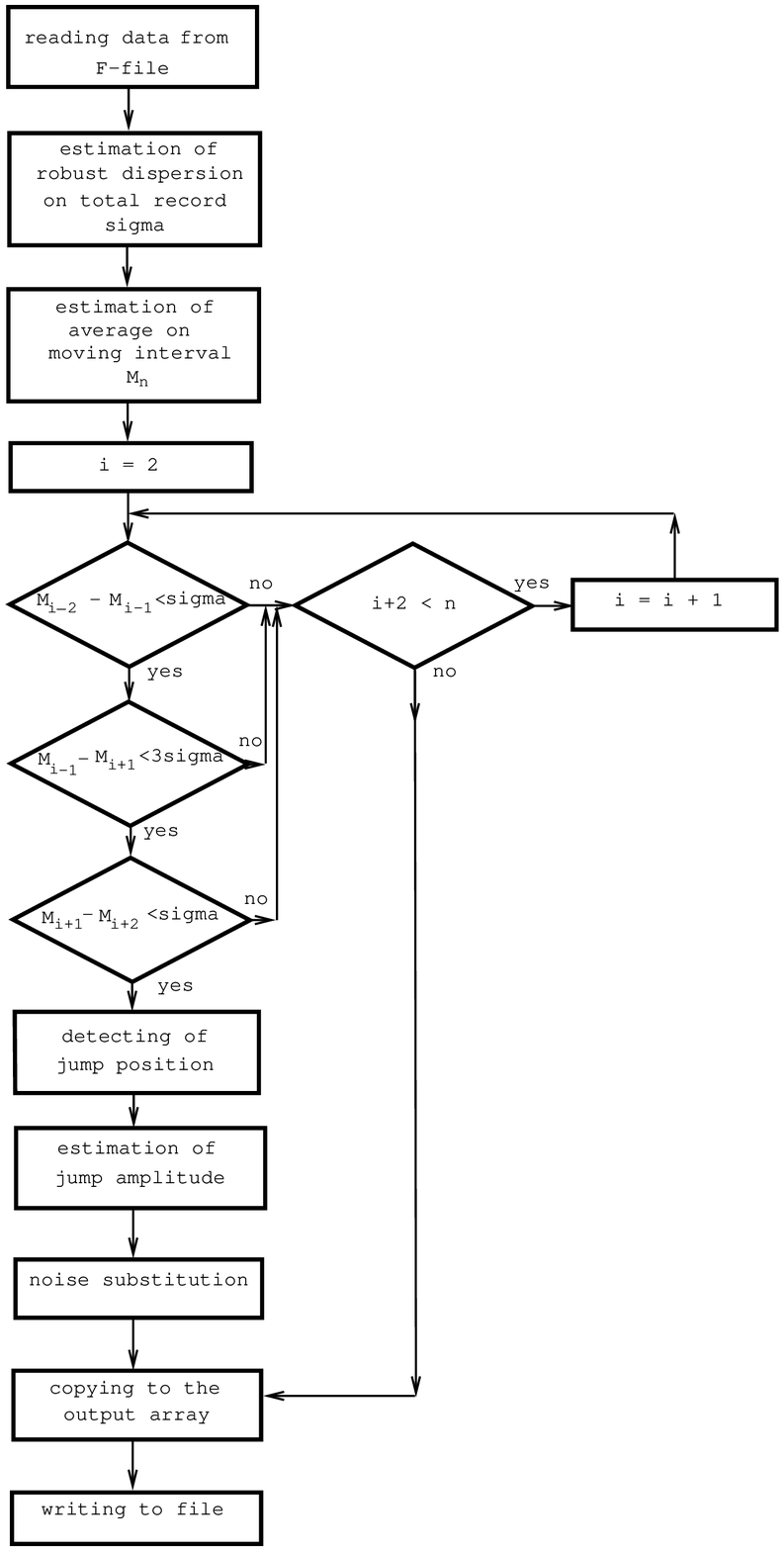,width=14cm,height=24cm}}
\caption{Block-diagram of search for and clipping  record jumps.}
\end{figure*}
\begin{figure*}
\centerline{\psfig{figure=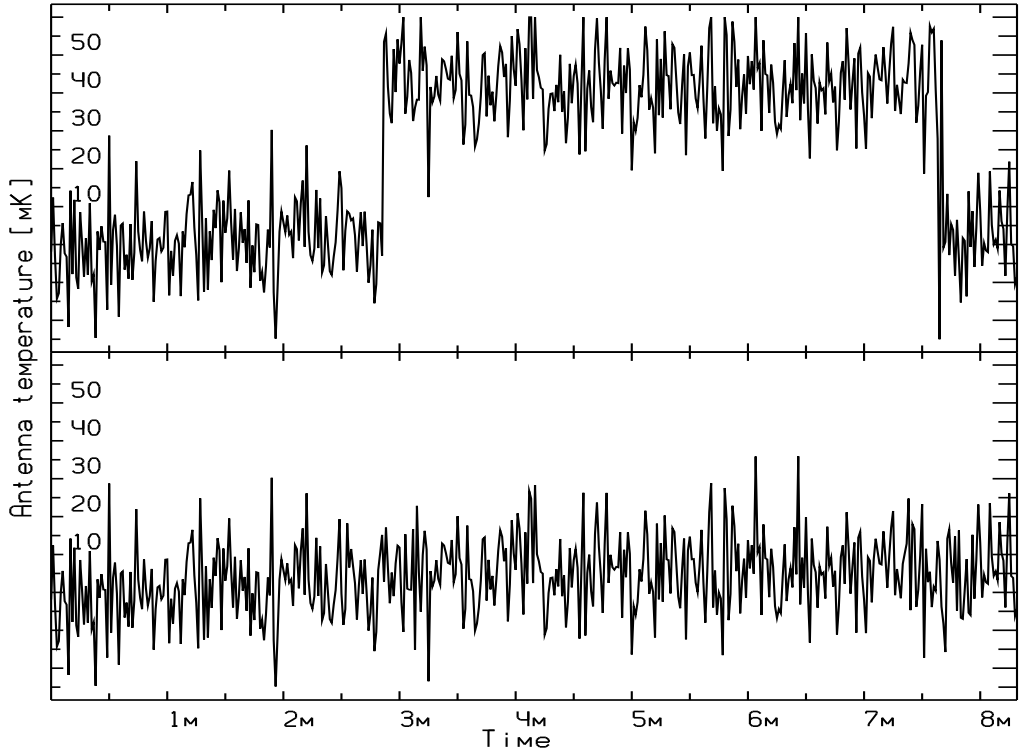,width=14cm,height=10cm,angle=-90}}
\caption{Results of removing jumps in a record.
The upper panel shows a record with a jump.
The lower panel shows a cleaned record.}
\end{figure*}

The algorithm is described as follows:
\begin{enumerate}
\item Reading data from the F file to the working array and estimation
of the robust dispersion $\sigma$ of the total record.
\item Filling the array $M_n$, where every element is an estimate of the
mean in the sliding window of a specified size. It is desirable that the
size of the window should be chosen not shorter than 5 pixels.
\item Location of the interval in which a jump occurred. Here we analyse
the array of the mean values estimated by a robust technique with the
aim of revealing the position of the abrupt increase of the mean
($>n \sigma$, by default $n$=3) value (gradient approach) by way of
comparing the values of four elements of the array $M$. The jump is
considered to be detected in the $i$-th interval provided the following
conditions are satisfied at the same time:
   $$
      | í_{i-2} - M_{i-1} | < \sigma,
   $$
   $$
      | M_{i-1} - M_{i+1} | < n \sigma,
   $$
   $$
      | M_{i+1} - M_{i+2} | < \sigma.
   $$

Thus the scheme of coincidence of the mean in two neighbouring
intervals at the edges and of uneven  transition in the central window
is realized.
\item Revision of the jump location (for instance, on the $k$-th pixel)
by means of comparison.
\item Decrease of the values of all pixels following the $k$-th pixel
by the value of the jump amplitude.
\item Decrease of all values in the array of the means in the sliding
window, which follow the current value of the mean.
\item If there is no limitation on the length of the record, a shift
along the record and jump to the condition of the check point are performed. In other
words, the cycle is completed.
\item Recording to the F file.
\end{enumerate}

In Fig.5 is shown a schematic diagram of the above-described algorithm
of searching for and removing the jump in the records. The results of
performance of the algorithm are demonstrated in Fig.6.

\section{Conclusions}
Through analysing statistical properties in the current interval of
the record the developed algorithms find and remove interferences of
different kinds quickly and effectively.

The procedure is involved as a command ``clip'' and included in the
standard data processing system FADPS at RATAN-600.

Provision is made for possible extension of the algorithm to
two-dimensional data and its application to removal of cosmic particle signals
on CCD images.
\begin{acknowledgements}
The authors owe debts of gratitude to N.F. Vojkhanskaya for valuable
comments in the course of preparing the manuscript.
\end{acknowledgements}

\end{document}